# GaAs quantum dots grown by droplet etching epitaxy as quantum light sources

Saimon Filipe Covre da Silva[a*], Gabriel Undeutsch[a], Barbara Lehner[a], Santanu Manna[a], Tobias M. Krieger[a], Marcus Reindl[a], Christian Schimpf[a], Rinaldo Trotta[b], and Armando Rastelli[a*]

[a]*Institute of Semiconductor and Solid State Physics, Johannes Kepler University, Linz, Altenberger Straße 69, 4040,Linz Austria*

[b]*Department of Physics, Sapienza University of Rome, Piazzale A. Moro 5, I-00185 Rome, Italy*

*corresponding authors:*

*Saimon F. C. da Silva: saimon.covre_da_silva@jku.at*

*Armando Rastelli:armando.rastelli@jku.at*

**ABSTRACT**

This paper presents an overview and perspectives on the epitaxial growth and optical properties of GaAs quantum dots obtained with the droplet etching method as high-quality sources of quantum light. We illustrate the recent achievements regarding the generation of single photons and polarization entangled photon pairs and the use of these sources in applications of central importance in quantum communication, such as entanglement swapping and quantum key distribution.

# I. INTRODUCTION

Non–classical states of light are the basis of various photonic quantum applications, like quantum communication and quantum key distribution (QKD),[1,2,3] quantum information processing[4] and quantum metrology.[5] Over the past several years, different approaches were developed to generate single photons[6] and entangled photon pairs.[7] Spontaneous parametric down conversion (SPDC) sources are the standard choice for the generation of heralded single photons and entangled photon pairs with a high degree of entanglement.[8] As an example of application, entanglement-based QKD[9] demands entangled pair sources with highest possible brightness and near-unity degree of entanglement, properties that current SPDC sources cannot provide at the same time. In fact, the stochastic photon generation process leads to an increase in multi-pair generation probability (and consequent degradation of entanglement) with increasing pump power and single-pair generation probability[10]. In this respect, quantum systems, capable of generating at most a single photon or a single photons pairs per excitation pulse are particularly attractive. Among solid-state emitters[11], semiconductor quantum dots (QDs)[12] are the only ones having demonstrated highly entangled photon pair emission, a process taking place via the biexciton-exciton (XX-X) decay cascade (for a recent review, see Ref[13]). In contrast to SPDC sources, QDs exhibit sub-Poissonian emission characteristics and therefore can keep an ultra-low multi-photon-pair emission probability independent of their brightness, see Refs.[10,14]

In general, the ideal source of single and entangled photons should satisfy three important requirements[12,15]: (i) The source should deliver photon (pairs) on demand in a pure quantum state. (ii) Subsequently emitted photons should be indistinguishable. (iii) The photons should be injected with the highest possible efficiency into the desired quantum channel. In addition, unavoidable transmission losses in the quantum channels (typically free-space[16] and optical fibers[17]) should be considered when choosing a photon source. A possible solution for this problem (which is currently limiting the practical transmission distances to some hundreds of kilometers) is based on the idea of quantum repeaters,[18] that may require the implementation of quantum teleportation[19–21] and entanglement swapping protocols,[22] where not only the entanglement fidelity is important but also the photon indistinguishability.[23–25]

In the last few years, GaAs QDs grown via the local droplet etching (LDE) method[26] on GaAs(001) substrates emerged as a promising platform for the generation of single photons and entangled photon pairs with a wavelength around 780 nm. While this wavelength is not compatible with existing optical-fiber infrastructure, it matches one of the atmospheric transmission windows (775-785 nm) favorable for satellite-based QKD[27] and also atomic transitions of Rubidium, which are being considered for the implementation of quantum memories[28] as elements of quantum repeaters[18] and the maximum efficiency of Silicon-based photodetectors. Different from QDs obtained via the more traditional Stranski-Krastanow (SK) method[29], these dots can be grown with almost no strain using the GaAs-AlGaAs material combination and are characterized by an improved ensemble homogeneity compared to commonly studied InGaAs QDs (see Ref.[30] and Ref. [26] for a discussion on the optical properties of QDs obtained with different growth methods). Under optimized growth conditions, highly symmetric GaAs QDs with ensemble-averaged excitonic fine structure splitting (FSS) comparable to the natural linewidth of the excitonic transition can be obtained.[31] As discussed below, a negligible FSS is highly desirable to reach near-unity degree of photon entanglement without resorting to lossy post-selection schemes.[32] Another special feature of GaAs QDs by LDE is that excitons are weakly confined, resulting in lifetimes for the neutral exciton state around 250 ps.[33] Such short lifetimes, combined with average FSS values below those typically reported for InGaAs QDs[30] are beneficial to achieve high degree of entanglement: fidelity values up to 0.978(5), the highest reported so far for any QD type were achieved in Ref.[34] In addition, progress in photonic integration has led to photon pair collection efficiencies around 0.65(4)[35] (single photon collection efficiencies of 0.85(3)). The Purcell enhancement in broadband cavities may further enable enhancing the photon indistinguishability of the XX-X cascade,[36] as this is predicted to depend on the lifetime ratio of the biexciton and exciton states. [37] The integration of GaAs QDs in charge-tunable devices has enabled blinking under resonant excitation to be suppressed and QD emission close to the Fourier limit, i.e. almost perfect photon indistinguishability, to be achieved.[38] Following material quality optimization and integration in charge-tunable devices, two-photon interference with a visibility of up to 93.0(8)% among photons emitted by two remote GaAs QDs has been recently reported.[39] This result sets the state-of-the-art of two-photon interference between photons from any two solid-state emitters. Finally, the very limited built-in strain and strain inhomogeneity compared to SK dots combined with the large dot dimensions promise enhanced coherence times for electron spins confined in GaAs QDs.[38]

Based on the above considerations, we believe that GaAs QDs produced by the LDE method will keep getting closer to the ideal source of single and polarization-entangled photon pairs (see the listed points (i), (ii) and (iii) in the second paragraph of the introduction).

## II. Epitaxial growth and properties of GaAs QDs

### A. Epitaxial growth of GaAs QDs by molecular beam epitaxy

In order to obtain QDs with optical properties suitable for demanding quantum-optical experiments and applications, several methods have been developed over the last decades. The most prominent rely on epitaxial growth of self-assembled nanostructures with techniques such as molecular beam epitaxy (MBE) and metalorganic vapor phase epitaxy. While most of the pioneering experiments in the field have relied on strained InGaAs QDs obtained by the SK mode[40], LDE has been developed more recently[41–43] and exploited to obtain almost strain-free and highly symmetric GaAs QDs with surface densities ideally suited for single QD applications. Initially the method was introduced by Wang et al.[41] for etching GaAs surfaces using Ga droplets and was extended later to different surfaces like AlGaAs using Al or In droplets,[42,43] as well as to InP [44] and GaSb substrates[45]. In contrast to the SK process, which takes advantage of lattice mismatch to grow 3D structures (e.g. Ge/Si[29,46], InAs/GaAs[40].), LDE does not rely on strain. Compared to conventional droplet epitaxy[47], which consists in exposing droplets to an As flux at low substrate temperatures, LDE is performed at higher temperatures, leading to improved crystal quality, as deduced from the generally narrower emission linewidths.[26] Figure 1(a) depicts the steps for obtaining GaAs/AlGaAs QDs via LDE. First, an AlGaAs layer is grown on a GaAs substrate that will act as a confinement barrier. Next Al droplets can be obtained by Al deposition in the absence of Arsenic. Al will react with the AlGaAs forming shallow depressions in the surface.[48] Specifically, the As concentration gradient across the interface between Al droplet and AlGaAs substrate leads to As diffusion into the droplet and consequent partial liquefaction (etching) of the AlGaAs material under the droplet.[48–50] The process is supposed to stop when the solubility limit of As inside the Al droplets is reached[48]. Although somewhat counterintuitive, further etching requires the exposure of the surface to an As flux (step 1 in Fig. 1a), which activates the diffusion of the Al (and Ga) atoms away from the droplet and their incorporation together with As (as recrystallized AlGaAs) in the regions around the nanoholes (step2 in Fig. 1a). The growth conditions, i.e. substrate temperature, deposition rate, and amount of deposited Al during step 1 determine the droplet density and size and are important to obtain symmetric nanoholes during the following step 2. The proper choice of As background to obtain highly symmetric nanoholes may depend on the details of the employed MBE chamber. For our MBE, very good results are obtained by gradually increasing the As beam equivalent pressure (BEP) from $3.0\times10^{-7}$ to $3.0\times10^{-6}$ and finally to $2.5\times10^{-5}$ mbar in steps of 1 min duration. A representative AFM image of nanoholes with a surface density of 0.2 $\mu m^2$ and highly symmetric shapes is shown in Fig.1 (b). The nanoholes are then filled with GaAs by depositing 1–4 nm of GaAs followed by a 1 min growth interruption to allow GaAs to diffuse into the nanoholes. A top AlGaAs barrier (step3) concludes the growth, leading to an AlGaAs/GaAs/AlGaAs type-I heterostructure with strain-free GaAs QDs. The depth of the nanoholes together with the choice of the GaAs amount during filling and the Al content of the barriers determine the average emission wavelength.[51]

## B. Optical properties of GaAs QDs

Due to the homogeneity of the nanohole size,[52] QDs in an ensemble emit in narrow spectral ranges, as, e.g, illustrated in the histogram of Fig. 1(c), which was constructed by measuring the energy of the neutral exciton (X) lines from micro-photoluminescence (μ-PL) spectra of 200 randomly selected QDs. Dots emitting in this wavelength range may be possibly interfaced with quantum memories based on optical transitions of Rb atoms,[53] such as the $^{87}$Rb D2 line (~780.24 nm).[54] Although this wavelength is not ideal for the standard fiber network (which operates mainly around 1550 nm), it is suitable for free-space communication[27] and could be further adapted via conversion techniques.[55]

Fig. 1(e) shows a representative PL spectrum of a GaAs QD obtained under above-bandgap continuous-wave laser excitation.[33] In addition to the isolated X emission, several lines are observed, which were partly ascribed to specific recombination channels by combining polarization, temperature and magnetic-field-dependent measurements with configuration-interaction calculations.[33] The comparison between experiment and theory also allowed us to conclude that excitons are rather weakly confined in the GaAs QDs and that the single-particle picture is poorly suited to describe the behavior of such QDs. Additional consequences of the weak confinement are the fast recombination times and – most probably – the slow relaxation upon quasi-resonant excitation via excited states.[56]

For the generation of entangled photon pairs via the XX-X cascade, a QD should be initialized in the |XX> state. To date, the best way to achieve state preparation with efficiencies of the order of 90%[24] is the two-photon excitation process (TPE), in which the pump laser energy is set to half of the energy separation between the |XX> state and the crystal ground state |0>, see inset of Fig. 1(f). Under these conditions, XX and X emission lines dominate the spectra and the laser stray light can be removed via notch filters,[30,57] as seen in Fig. 1(f). To have highly entangled photon pairs, it is also desirable to have no FSS for the intermediate X levels.[58] The high in-plane symmetry achievable for GaAs QDs reflects in average FSS values as small as (2.5±1.3) μeV, as illustrated in the histogram of Fig. 1(d). This value is comparable to the Fourier transform limited linewidth of the X emission ($\frac{\hbar}{\sim 250\ ps}$~2.3 μeV) and probably the lowest average FSS value reported to date. This result makes it relatively easy to find "entanglement-ready" QDs, in contrast to other QD systems.

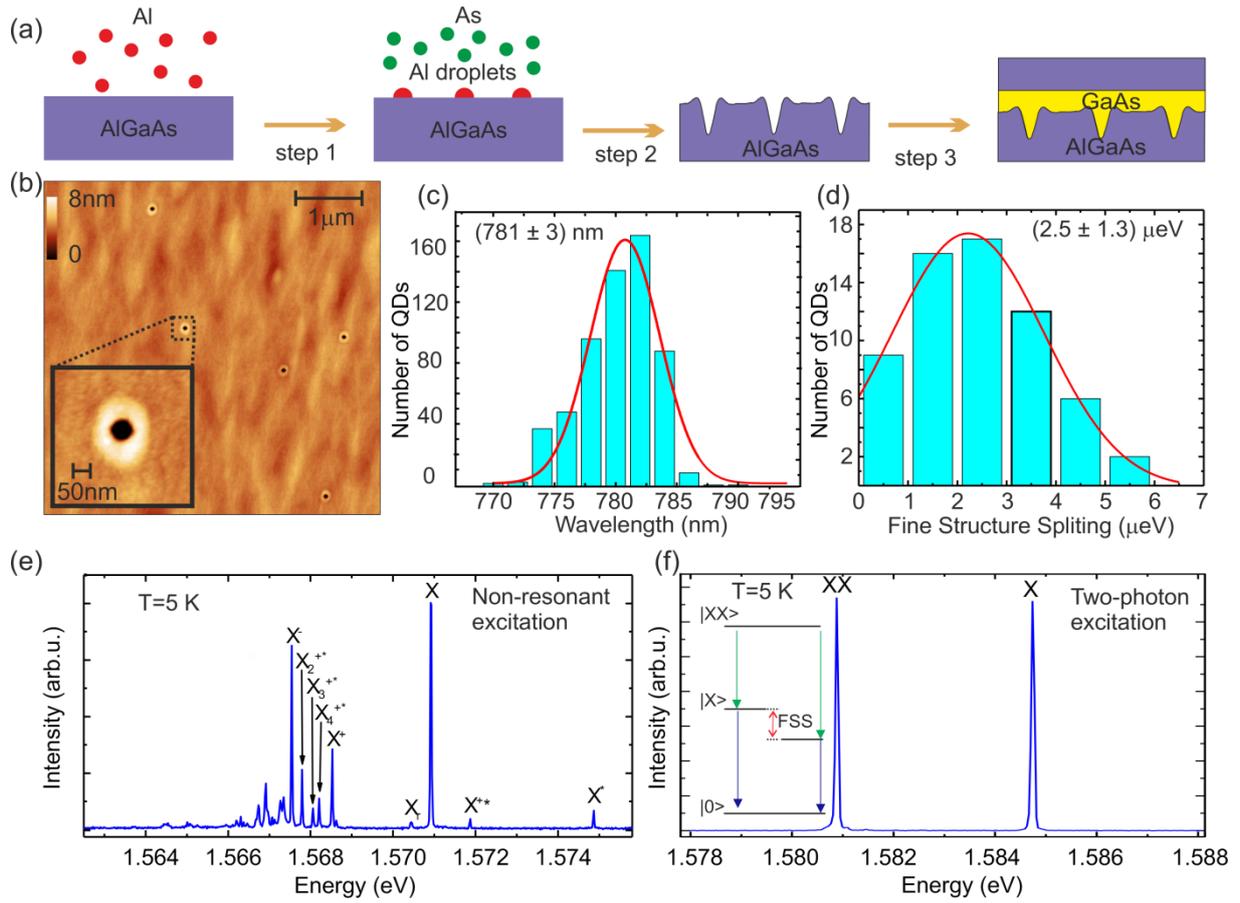

FIG. 1 (a) Illustration of the local droplet etching technique to grow highly symmetric QDs by GaAs filling of nanoholes obtained by Al-etching on an AlGaAs(001) surface. Step 1 consists of Al deposition in absence of As flux and consequent droplet formation. During Step 2 a growth interruption under As flow promotes etching and nanohole formation. Step 3 consists in nanohole filling and overgrowth. More details are provided in the text. (b) Atomic force microscopy image of LDE nanoholes obtained by deposition of 0.5 ML of Al equivalent monolayers (ML) at a substrate temperature of 600 °C and a rate of 0.5ML/s followed by annealing under As flux. (c) Distribution of dot emission wavelengths for QDs obtained by overgrowing the nanoholes with 2.0 nm of GaAs, the curve is a Gaussian fit. (d) Histogram obtained by measuring the excitonic FSS of 60 randomly chosen QDs. (e) The panel shows a typical photoluminescence spectrum of a single QD obtained by above-band excitation. The neutral exciton, X around 1.571 eV and other recombination lines are visible, as discussed in Ref. [33]. (f) Emission spectrum under pulsed resonant two-photon excitation (TPE) of the biexciton state, the inset show the involved states, i.e. the neutral biexciton |XX>, the exciton |X> and the crystal ground state |0>.

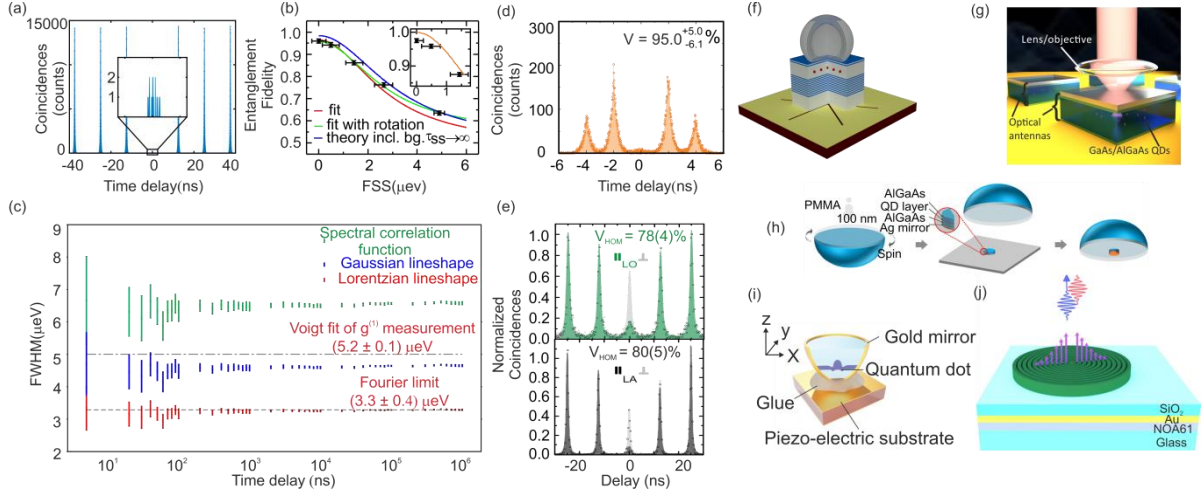

FIG. 2. (a) Autocorrelation histogram of the biexciton signal from a GaAs QD under TPE. The magnified portion shows the events around zero-time delay. Reproduced with permission from Schweickert et al., Appl. Phys. Lett. **112**, 093106 (2018), Copyright 2018 AIP Publishing (Ref.[59]). (b) Measured data (symbols) and modeling results (lines) for the entanglement fidelity versus FSS for a GaAs QD, figure reproduced with permission from Huber et al., Phys. Rev. Lett. **121**, 033902 (2018), Copyright 2018 AIP Publishing (Ref[34]). (c) Results of PCFS measurements on an exciton confined in a GaAs QD. The FWHM of the spectral correlation function (green) is plotted as a function of the time delay between photon detection events. To compare the results with the Fourier limit and the FWHM obtained from coherence ($g^{(1)}$) measurements, two lineshapes are assumed: Gaussian (blue) and Lorentzian (red). (d) HOM measurements for photons emitted within 2 ns by a trion under resonant fluorescence conditions. Figure reproduced with permission from Nano Lett. **19**, 2404 (2019) (Ref[70]). (e) HOM measurements for photons emitted within 12 ns by a neutral exciton under LO-phonon (green) and LA-phonon (black) excitation. Reproduced with permission from Reindl et al., Phys. Rev. B. **100**,155420 (2019), Copyright 2018 AIP Publishing (Ref[56]). (f-j) Different optical designs implemented so far to increase the light extraction efficiency for GaAs/AlGaAs QDs grown by LDE: (f) A planar distributed Bragg reflector cavity containing GaAs QDs with a solid immersion lens on the top, (g) planar metal-semiconductor-metal Yagi-Uda antenna containing GaAs QDs. (h) Planar antenna where a GaP solid immersion lens is used for coupling the evanescent light from the semiconductor layer using a PMMA gap layer. Reproduced from Chen et al., Nat. Comm. **9**, 2994 (2018), (CC BY) license (Ref.[60]). (i) parabolic mirror cavity structure. Used under permission from Lettner et al. ACS Photonics 7, 29 (2020) with CC-BY licence Ref[61]. (j) Deterministically fabricated CBR cavity structure (used under permission from Nat. Nanotech. **14**, 586 (2019) Ref[35]).

## C. Multi-photon emission probability:

The probability of a source to generate exactly one photon (or photon pair in the case of the XX-X cascade) in response to a trigger pulse is described as single photon purity. The multi-photon emission probability is quantified by the measurement of the second-order correlation function $g^{(2)}(0)$, using a standard Hanbury-Brown and Twiss (HBT) setup. An example of a HBT measurement performed by Schweickert et al.[59] on a GaAs QD under TPE can be seen in Fig. 2(a). By carefully suppressing the laser background and by using single photon detectors with ultra-low background, an unprecedentedly low $g^{(2)}(0) = (7.5 \pm 1.6)\times 10^{-5}$ was measured without any background subtraction or temporal post-selection and – most importantly – at maximum intensity (π-pulse conditions). Due to the finite pulse duration (typically about 10 ps), two-level systems suffer from a finite re-excitation probability, which causes undesired multi-photon emission. By using the TPE scheme re-excitation is strongly suppressed because the QD can be re-excited only when both the biexciton and exciton have recombined[62] (the average lifetime of the cascade, ~370 ps, is much longer than the biexciton lifetime, ~120 ps, and thus of the laser pulse). It is interesting to note that the observed value of $g^{(2)}(0)$ for photons emitted by a single GaAs QD is comparable to the lowest reported so far for a single ion, $g^{(2)}(0) = (8.1 \pm 2.3) \times 10^{-5}$, see Ref[63] with the advantages that a QD does not require traps and can be driven at higher repetition rates due to the typically shorter lifetimes compared to atomic transitions.

### D. Polarization Entanglement:

Due to the optical selection rules, the XX-X cascade in a QD with vanishing FSS should produce two photons with polarization in the $|\phi^+\rangle = \frac{1}{\sqrt{2}}(|HH\rangle + |VV\rangle)$ Bell state.[58] Several degrading effects lead to deviations from this ideal situation.[64] The probability of finding the two-photons in the ideal state is given by the entanglement fidelity.[65]. To bring the FSS below the detection limit and to study the effect of increasing FSS on the entanglement fidelity, Huber et al[34] used a micro-machined piezoelectric actuator[66] to fine tune the in-plane stress of a QD and surrounding matrix. For vanishing FSS, an entanglement fidelity as high as 0.978(5) was measured, without resorting to any post selection techniques. To date, this is the highest fidelity value reported for any QD system. The evolution of the fidelity for increasing FSS is shown in Fig. 2(b) (black points) together with a fit using the spin-scattering model presented in Ref[32] (red curve) and by taking into account only a time-dependent phase introduced by the FSS (blue curve). The inset figure (orange curve) shows the evolution of the fidelity versus the FSS for background-free entangled photons and no spin dephasing. From this work, it was concluded that moderate spin-scattering processes are still present, which could be alleviated by further reducing the recombination time via the Purcell effect.[35]

### E. Emission linewidth and indistinguishability:

Ideally, the recombination of an exciton should lead to a Lorentzian emission line with a FWHM given by $\hbar/T_1$, where $T_1$ is the transition lifetime (Fourier limit). In reality, several mechanisms inherent to the solid-state environment lead to line broadening and decreased photon indistinguishability. The most studied are pure dephasing,[67] charge and spin noise.[68] The latter ones typically occur on time scales much longer than $T_1$. In the case of GaAs QDs, the noise dynamics have been recently investigated by photon correlation Fourier spectroscopy (PCFS).[69] Fig. 2(c) shows the results of PCFS measurements on an X confined in a GaAs QD under TPE at time scales ranging from few nanoseconds to milliseconds. Different from the noisy QDs studied in Ref.[69], the FWHM of the spectral correlation function stays constant over the whole studied timescales. Although PCFS measurements do not give access to the true emission lineshape, we compute the linewidth for a Lorentzian shape (expected at short time scales) and a Gaussian shape (expected in the case of large noise contribution) and compare them with results of $g^{(1)}(\tau)$ coherence measurements obtained via Michelson interferometry and the Fourier-limited linewidth (3.3 μeV for the chosen QD). From the $g^{(1)}(\tau)$ measurements (providing information on the linewidth averaged over timescales of a few minutes), we extract a linewidth which is 1.6 times larger than the Fourier limit a significant improvement compared to former works.[70,71] Even better results, down to 1.04 times the Fourier-limit have been obtained by Zhai et al. under cw resonant fluorescence conditions in charge-tunable devices employing a p-i-n diode structures.[38]

The photon indistinguishability can be quantified by Hong-Ou-Mandel (HOM) interference measurements.[12,72] Fig. 2(d) shows a coincidence histogram recorded in HOM measurement performed with a GaAs/AlGaAs QDs by Schöll et al.[70] The QD was excited with double laser pulses in resonance with the positive trion transition and separated by a delay of 2 ns. HOM interference visibilities of V = 95(5)% were obtained without any Purcell enhancement. In another work, Reindl et al.[56] studied also the phonon-assisted excitation of neutral excitons in GaAs. Fig. 2(e) depicts representative results using LO-phonon (green) and LA-phonon (black) assisted excitation for time delays of 12 ns with the visibilities of $V_{X_{LO}}^{HOM}$ = 78(4)% and $V_{X_{LA}}^{HOM}$ = 80(5)% respectively. The visibility degradation compared to shorter delays are attributed again to residual charge noise in the sample used for that study.[69,73] Very recently, Zhai et al. have demonstrated that the Fourier-limited emission of GaAs QDs in p-i-n diodes corresponds to perfect photon indistinguishability under pulsed excitation for pulse separation exceeding 1 μs.[39] In addition, two-photon interference with visibilities exceeding 90% between photons emitted by remote GaAs QDs were demonstrated. These results make GaAs QDs the highest quality QD emitters to date.

Finally we note that the indistinguishability of photons emitted in the XX-X cascade is intrinsically limited by the time correlations between the two photons.[37] This issue may be alleviated by resorting to differential Purcell enhancement and spectral filtering, as discussed in detail in Ref.[36]

**F. Brightness:**

The brightness of a QD source can be defined as the probability of collecting a single photon (or a photon pair) with the external optics (first lens) for each excitation pulse. In the case of a matrix composed of GaAs and assuming unity quantum efficiency and preparation fidelity, only 2 % of the light can be coupled at maximum into free space, the rest of the light will be lost due total internal reflection.[74] In the last several years, different approaches were developed to overcome this problem, such as solid immersion lenses (SILs),[75] planar Fabry-Perot cavities based on distributed Bragg reflectors (DBR),[76] micropillars,[77] photonic trumpets,[78] circular

Bragg grating structures,[79] and open cavities.[80] Here we discuss only the photonic structures used so far for GaAs QDs.

The first design sketched in Fig. 2(f) and used several times,[34,56,59,69,70] because of its comparatively simple growth and processing, is a DBR structure consisting of a lambda cavity, sandwiched between 2 top and 9 bottom pairs of $Al_{0.95}Ga_{0.05}As/Al_{0.20}Ga_{0.80}As$ layers. For example, in the work of Schöll et al,[70] a zirconia SIL was added, resulting in an extraction efficiency (first lens) of 20(3) % (for a numerical aperture of 0.81). By interrupting the substrate rotation during MBE growth of the top cavity layer, a gradient in the mode position can be induced that make the approach tolerant to fluctuations in the growth.[59]

Fig. 2 (g) shows the design of a planar antenna used by Huang et al.[81] The antenna consists of a metal reflector and a semi-transparent director, with a thin oxide layer between metal and semiconductors layer. An extraction efficiency of ~19 % (for a numerical aperture (NA) of 0.81) and a low multiphoton emission probability of 0.006(5) were obtained. Although higher efficiencies are predicted for thinner designs, line broadening effects caused by surface charge fluctuations make the implementation difficult. The second important point is that the antenna has a broad bandwidth of ~30 nm, which is tolerant to fabrication imperfections.[82,83]

In another work, Fig. 2(h), Chen et al.[60] used frustrated total internal reflection[84] to build up a broadband optical antenna. By using an accurately sized gap (filled with PMMA) between the QD structure and a GaP hemi-spherical lens, a photon extraction efficiency of 0.65(4) was obtained. Under resonant TPE excitation, it was demonstrated that the device can be used as highly-efficient entangled-photon source with entanglement fidelity of 0.90 and a $g^{(2)}(0)$ of 0.002(2).

Fig. 2 (i) shows the design proposed by Lettner et al.[61], which ideally consists of a deterministically positioned GaAs QD in a parabolic structure with a gold back mirror, using a fabrication method based on photoresist reflow. The idea is to use the parabolic mirror to redirect the QD emission towards the collection optics. The fabrication process of this kind of structure is more complex than those of a DBR or a planar antenna, since a precise positioning of the QD and a cutting-edge technology for the fabrication of the parabolic mirror is needed. The parabolic microcavities were integrated on top of piezoelectric actuators, providing the possibility to tune the QD emission. The obtained extraction efficiency was 12 %, six times less than the simulated value, which was attributed to the random QD lateral position in the cavity and fabrication imperfections. A major advantage of the parabolic microcavity design is that the diameter is comparable to the core diameter of standard optical fibers, making a direct integration of a fiber on the top of the device possible.

The currently most promising structure for entangled photon sources is shown in Fig. 2(j). It consists of a circular Bragg resonator on top of a highly efficient broadband reflector (CBR-HBR). With this device, a single photon extraction efficiency of 0.85(3) and a pair collection efficiency up to 0.65(4) were demonstrated using GaAs QDs[35]. The lifetime of the exciton was shortened from 210 ps to 60 ps, which corresponds to a Purcell factor of 3.5. Cross-correlation measurements between X and XX under π-pulse excitation, resulted in an entanglement fidelity of 0.88(2) for a QD with a FSS about 4.8 µeV. Besides the photonic environment, other factors affect the brightness of a QD source. The most important ones are (i) the state preparation efficiency (the probability of creating a certain exciton state upon an excitation pulse), (ii) blinking, i.e. suppressed excitation/recombination and (iii) non-radiative recombination. For GaAs QDs, (i) has been especially studied under TPE and phonon-assisted TPE[24,85], and (ii) under different resonant excitation conditions.[56,69,86] The main source of blinking, which severely reduces the source brightness, are random charge fluctuations, which can be suppressed by embedding the QDs in diode structures.[38,39]

## III. APPLICATIONS OF QD-EMITTED ENTANGLED PHOTONS

GaAs QDs obtained with the LDE method were used to implement quantum teleportation of the polarization state of a photon emitted by a QD to another photon emitted by the same QD and belonging to an entangled photon pair.[23] Fig. 3(a) shows the third-order correlation function (upper panels), from which the teleportation fidelities (lower panels) for different input polarization states are calculated. The teleportation fidelity averaged over all three input states is 0.75(2).

The next logical step after teleportation is "entanglement swapping", which consists in establishing entanglement between two photons belonging to two initially uncorrelated entangled-photon-pairs. Entanglement swapping relies on fourfold coincidence events whose rate has a fourth-power dependence on the light extraction efficiency.[24] The first realizations of entanglement swapping using a solid state quantum emitter were reported by Basso Basset et al.[24] and M. Zopf et al.[87], again using GaAs QDs. Fig. 3(b) depicts the measurement results (swapping fidelity vs FSS and HOM interference visibility) together with a theoretical model from Ref.[24] which takes two main parameters into account: the initial degree of entanglement of the two photon pairs (linked to the FSS) and the indistinguishability of the photons used for the BSM. According to the best values for entanglement fidelity[34] and indistinguishability[35] showed in literature for GaAs LDE-QDs, the model shows that a swapping fidelity and concurrence of 0.84 and 0.67 can be obtained, values that can violate Bell´s inequality.[88]

In another work Zopf et al. [87] used a GaAs QD embedded in a planar antenna Fig. 2(h) to show the violation of the CHSH and Bell inequalities with a swapping fidelity of 0.81(4), though with temporal post-selection.

The high entanglement fidelity of photon pairs emitted by GaAs QDs makes these sources very well suited for entanglement-based QKD, as recently demonstrated in Refs. [14,89] As an illustration, Fig. 3(c) displays the encryption of a bitmap image performed in a QKD experiment done by Schimpf et al.[14] using a GaAs LDE-QD and the BBM92[90,91] protocol. The key generation was performed between two buildings, connected by an optical fiber of 350 m, which resulted in a raw key rate of 135 bits/s and a qubit error rate of 1.91 %. In parallel, Basso Basset et al. [89] used a similar structure but a modified Ekert QKD protocol using a 250 m long free-space quantum channel.

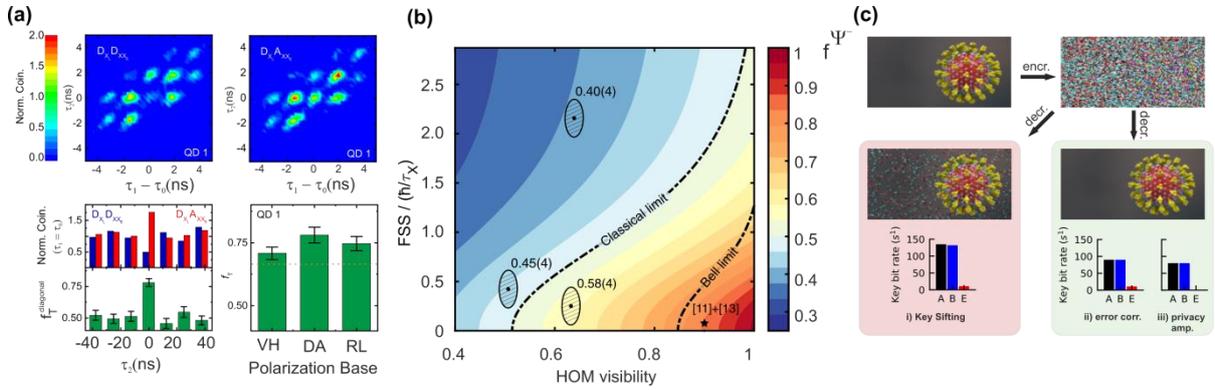

FIG. 3 (a) Results showing the third-order correlations functions for the photons measured in a teleportation experiment using trains of entangled-photon pairs emitted by a GaAs LDE-QD. Adapted with permission from Reindl et al. Sci. Adv. **4**, eaau1255 (2018) under CC BY license (Ref.[23]). (b) Experimental data and theoretical results showing the expected swapping fidelity as function of the HOM interference visibility and excitonic FSS for the entanglement swapping experiment. Reproduced with permission from Basset et al., Phys. Rev. Lett. **123**, 160501 (2019), Copyright AIP Publishing (Ref.[24]). c) Bitmap image encryption and decryption using a quantum key generated using the BBM92 protocol employing a GaAs QD as the entangled photon source. Adapted with permission from Schimpf et al. Sci. Adv. **7**, eabe8905 (2021) under CC BY license (Ref.[14]).

## IV. Summary and Outlook

In this paper we have reviewed the main properties of GaAs QDs obtained by a self-assembly process relying on nanohole formation on an AlGaAs layer via the local droplet etching (LDE) method[41] and GaAs/AlGaAs overgrowth, similar to the architecture first proposed in Ref.[92] These QDs have recently outperformed conventional InGaAs/GaAs QDs (and any other solid-state quantum emitter) as sources of single and polarization-entangled photon pairs according to three key figures of merit: 1) they have shown the lowest multi-photon emission probability reported so far in the literature,[59] [$g^{(2)}(0) = 0.00075(16)$] under two-photon-excitation (TPE) conditions without resorting to background subtraction or temporal post-selection; most importantly, this value is independent from the source brightness, which is in stark contrast to SPDC-based sources,[93] 2) they have shown the highest degree of polarization entanglement [fidelity to the maximally entangled $|\phi^+\rangle$ Bell state of 0.978(5) and concurrence of 0.97(1)],[34] thanks to the ultralow ensemble average FSS (see Fig. 1d) and its further elimination via strain-tuning[66], the TPE method, and the intrinsically short lifetimes stemming from weak confinement[56]; 3) they have led to the highest two-photon (Hong-Ou-Mandel) interference visibility among photons emitted by two remote QDs [93.0(8)%] embedded in charge-tunable devices and driven under resonant optical excitation; it is important to note that this result was obtained without any spectral filtering nor resorting to the Purcell enhancement and is consequence of the high material quality and QD homogeneity reachable with the GaAs/AlGaAs platform.

In addition, the use of circular Bragg grating cavities has allowed the generation of entangled photon pairs with a pair collection probability of 0.65(4) and Purcell-enhanced emission rates.[35] Recent work on GaAs QDs into p-i-n diodes shows completely suppressed blinking under resonant single-photon[38] and also under TPE,[94] which is required for entangled-photon generation.

Although the topic is beyond the scopes of this review, we are optimistic that GaAs QDs will outperform InGaAs QDs also as hosts of stationary qubits (electron spins and nuclear spin ensembles)[95] thanks to the reduced strain and relatively large size, which reduce the impact of nuclear spin fluctuations. In turn, by creating efficient spin-photon interfaces,[80,96,97] the nuclear spins of the GaAs QDs could possibly be used as quantum memories.[98] This would open up the thrilling possibility of using GaAs QDs to build-up complete and scalable quantum repeater nodes connected by entanglement swapping operations.[36]

We now provide our view on future developments in the field. First of all, we believe that material optimization should continue across different laboratories and standardized material assessment protocols should be developed to guide this optimization. The target is to achieve widespread availability of highest quality materials across the scientific and engineering community and to explore the physical limits of the QD hardware, reduce the QD-to-QD variabilities and find solutions to cope with residual imperfections by understanding the underlying physics. Most of the detailed investigations so far have focused on QDs with emission wavelengths around 780 nm and produced by GaAs filling of AlGaAs nanoholes with an Al fraction between about 20% and 40%. It is not yet known if further improvement (in entanglement quality, lifetimes, photon indistinguishability, photon generation efficiency) can be achieved by using different recipes for nanohole formation and/or by reducing/increasing the dot size or by working with different barrier and filling material. Further effort has to be put in the engineering of the photonic environment of the QDs, which is necessary to funnel efficiently photons out of (or into) the QDs. We believe that a scalable QD architecture will require structures enabling several degrees of freedom to be tuned independently to achieve at the same time wavelength matching among different emitters, charge control, and FSS elimination.[99] Therefore, the photonic structures should be compatible with such tuning methods, as discussed in the outlook of Ref.[36] A promising approach is represented by the GaAs/AlGaAs QDs in a circular Bragg grating cavities [77,79,100,101] integrated on a microprocessed piezoelectric actuator[66]. Such device could potentially provide entangled photon pairs with near-unity entanglement and pair-collection efficiency well above the state of the art. Electric control would be used to suppress blinking and an asymmetric Purcell enhancement could also provide an improvement in the indistinguishability of photons emitted by the XX-X cascade.[102] This device could be immediately used to implement entanglement-swapping among remote emitters, an important milestone towards a quantum repeater.

**ACKNOWLEDGMENTS**

This work has received funding from the Austrian Science Fund (FWF): FG 5, P 29603, P 30459, I 4380, I 4320, and I 3762, the Linz Institute of Technology (LIT) and the LIT Secure and Correct Systems Lab funded by the state of Upper Austria and the European Union's Horizon 2020 research and innovation program under Grant Agreement Nos. 899814 (Qurope), 654384 (ASCENT+), and 679183 (SPQRel)

**DATA AVAILABILITY**

Data is available upon request.